\documentstyle[aps,amsmath,epsf,psfig,multicol]{revtex}
\begin{document}
\draft
\author{Jeanman Sur$^*$, Andrea L.  Bertozzi$^{**,*}$, and Robert P. Behringer$^*$}
\address{*Department of Physics and Center for
Nonlinear and Complex Systems,\\
**Department of Mathematics,\\ Duke University, Durham, NC 27708}
\title{Reverse undercompressive shock structures in driven thin film flow}
\date{\today}
\maketitle

\begin{abstract}

We show experimental evidence of a new structure involving an
undercompressive and reverse undercompressive shock 
for draining films driven by a surface tension gradient against
gravity.
The reverse undercompressive shock is unstable to transverse
perturbations while the leading undercompressive shock is stable.
Depending on the pinch-off film thickness, as controlled by the meniscus,
either a trailing rarefaction wave
or a compressive shock separates from the reverse undercompressive shock.
\end{abstract}
\pacs{PACS numbers: 68.15.+e,68.45.Gd,03.40.Gc,47.20.Ma}
\begin{multicols}{2}

We consider a thin fluid film wetting a solid substrate and driven
by a thermal gradient against gravity.  This system exhibits a range
of phenomena including two kinds of shocks (Lax and undercompressive)
and a fingering instability that forms from a flat interface.  A key
issue that we explore is the role played by the meniscus at the lower
boundary.

Insight is provided from two previous studies.  In experiments by
Ludviksson and Lightfoot\cite{LL71} (LL), a partially immersed
substrate was pulled from a liquid bath and allowed to drain.  A
temperature gradient drove the film up the substrate against gravity;
no fingering instability was reported.  Cazabat et al. (Cal) created
very thin capillary driven films\cite{CHTC90,CC93} by leaving the
lower end of the substrate in a liquid reservoir; the resulting
meniscus controlled the climbing film thickness.  A strong Marangoni
stress created a capillary ridge, resulting in a fingering instability
similar to that in gravitationally driven films \cite{hupp}.

In both types of experiments, the film thickness is a key parameter.
In the Cal configuration, inclining the substrate changes the
curvature of the meniscus, resulting in a range of film
thicknesses\cite{FCQ96}.  Cazabat et al.\cite{BMFC98,SC1-00} show that
the bulk film thickness can increase more than an order of magnitude
with decreasing inclination angle $\alpha$ from the horizontal.  As
$\alpha$ varies, a transition occurs from a stable flat front to a
front that is unstable to fingering.

This transition reflects a basic change in the dynamics of the film:
very thin film fronts show a single shock profile that is unstable to
fingering \cite{KT97} while thicker films exhibit an
undercompressive-compressive shock pair \cite{BMFC98,BMS99} in which
the leading undercompressive (UC) front is stable \cite{KT97,BMSZ}.
The experiments in \cite{BMFC98,SC1-00} are, to our knowledge, the
first examples of naturally occurring UC shocks for systems subject to
a scalar conservation law.

In this Letter
we present experimental evidence of a new structure involving a
pair of UC shocks.  This was recently proposed by M\"unch \cite{M02}
to explain dynamics mentioned in the original LL work \cite{LL71}.
Using a lubrication model for the meniscus region, he computes film
dynamics for an initial condition corresponding to the LL experiment.
In this model, the meniscus region pinches off a body of fluid that
travels up the substrate almost as a solitary wave.  This structure
contains leading and trailing fronts, both undercompressive.  The
trailing undercompressive front has never been documented
in experiments, although its existence was rigorously
proved\cite{BS00}.  M\"unch calls this
trailing front a `reverse undercompressive' (RUC) shock because it involves a
thicker film receding upward from a thinner film, which is the reverse
of the type found in \cite{BMFC98,SC1-00}, in which a UC shock
describes a thicker film advancing into a thinner region.

We compare theory for the film dynamics to our experiments. 
M\"unch uses a full meniscus model to describe the film and
meniscus.  In his simulations, 
the meniscus quickly pinches off the draining film, and
later acts as a boundary condition for the climbing film
structure.  In the early stages of the experiment, we clearly observe 
this pinch-off process. 
Here we model the dynamics of the draining film
using a well-known lubrication approximation with a `depth averaged'
velocity \cite{BMFC98}
\begin{equation}
 \vec V = (\frac{\tau h}{2\eta} - \frac{\rho g h^2\sin
 \alpha}{3\eta})\vec e_x
+\frac{\gamma h^2 \nabla^3 h}{3\eta}.
\label{model}
\end{equation}
We model the meniscus as a boundary condition that we discuss later.
In the above, $\gamma$ denotes the surface tension,
 $\tau = d\gamma/dx$ denotes the surface tension gradient, $\alpha$ the angle
of inclination (from the horizontal) of the plane, $g$ the
gravitational constant.  Also, $x$ is along the direction of the flow.
The coefficient of $\vec e_x$ in Eq.~(\ref{model}) represents
convection due to the surface tension gradient and the component of
gravity tangent to the surface.  The component of gravity normal
to the surface has a negligible effect.  We couple Eq.~(\ref{model})
with mass conservation, $h_t + \nabla\cdot(h \vec V) = 0$.  To
understand the shock dynamics, we ignore perturbations transverse to
$\vec e_x$ and consider solutions $h$ depending only on $x$ and $t$:
$h_t + (f(h))_x = -((\gamma/3\eta) h^3 h_{xxx})_x$.  The flux
satisfies 
$ f(h) = ({\tau h^2}/{2} - {\sin\alpha \rho g h^3}/{3})/\eta$.  
We rescale to dimensionless units as in \cite{BMFC98}: $h=H\hat h$,
$\quad x = \hat x l$, $\quad \text{and}\quad t= T\hat t,$ where 
$H = \frac{3\tau}{2\sin \alpha\rho g},$ 
$l = (\frac{2\gamma}{3\tau H^2})^{1/3} = (\frac{3\gamma\tau}{2\rho^2
g^2 sin^2\alpha})^{1/3},$ 
and $T=2\frac{\eta}{\tau^2}(\frac{4}{9}\tau\gamma\rho g sin\alpha)^{1/3}.$
Dropping the $\hat {\phantom{x}}$ gives the dimensionless equation
\begin{equation}
h_t + (h^2-h^3)_x = -(h^3h_{xxx})_x.
\label{non-dimen}
\end{equation}

\begin{figure}[h] 
\center{\parbox{3.2in}{\psfig{file=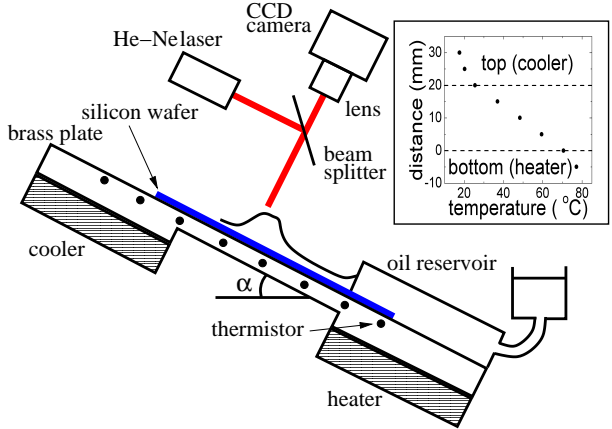,width=3.2in}}} 
\caption{The schematic diagram of the experimental setup. A typical temperature gradient 
is shown in the inset.}
\label{fig:apparatus} 
\end{figure}

The experiments are carried out as follows.  An oxidized silicon wafer
is partially dipped into a reservoir of silicone oil (PDMS, $\eta$=100
cSt and $\gamma$=0.0209 N/m at 25$^{o}$C) attached to a brass plate.
The wafer is pulled out and clamped to the plate
(Fig.~\ref{fig:apparatus}).  We create a uniform temperature gradient
along the plate by heating at the bottom with a foil heater and
cooling at the top with a circulating bath cooler.  We monitor the
temperature along the plate via thermistors.  A typical gradient,
which is constant during an experiment, is shown in the inset of
Fig. 1.  The distance between the heater and cooler is 20 mm, and the
width of the brass plate is 152 mm (effectively infinite width in the
transverse direction).

We measure the thickness of the film by both standard interferometric
techniques and by a novel variation of this technique.  The primary
technique uses collimated light from a He-Ne laser (wavelength 632.8
nm) incident normally on the film.  We record images with a CCD camera
and a frame grabber.  Additionally, at the end of a run, we shine an
infrared laser (830 nm, 100 mW) at a point on the film.  The film near
the laser spot is reduced to essentially zero thickness due to the
combination of local heating and the induced surface tension gradient.
We then measure the thickness at other points on the film by counting
the interference fringes from the reference position (One fringe
$=0.226 \mu$~m)\cite{SC1-00}.  We have less precision
determining the absolute thickness of very thick regions, although the
relative error between nearby fringes is small.  This is reflected in
the larger absolute error bar shown in the top right corner of Figs. 5
and 6.

\begin{figure}[h] 
\center{\parbox{3.2in}{\psfig{file=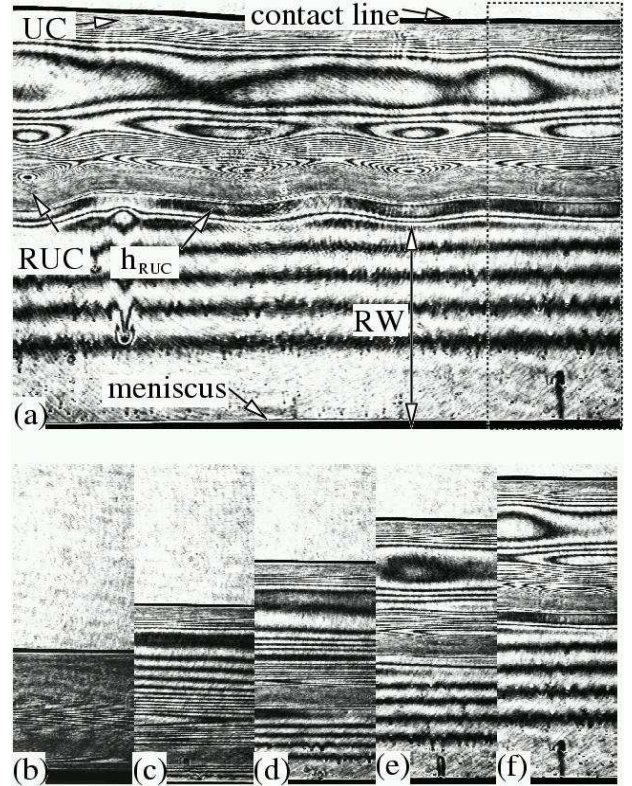,width=3.2in}}} 
\caption{(a) Interference fringes at t = 3200 sec.
for $\alpha=85^{\circ}$ and $\tau$ = 0.11 Pa.
The vertical and the horizontal size are 14.4 mm and 19.2 mm respectively.
Below are sections (as indicated by the dashed line in (a))
at times (b) 0 sec, (c) 800 sec, 
(d) 1600 sec, (e) 2400 sec, and (f) 3200 sec. 
}
\label{fig:profile} 
\end{figure}

In the early stage of the film evolution, a stationary pinched-off
portion appears in the meniscus, while the contact line climbs.
Later, a RUC shock moves up from the pinched-off portion
and a broadening RW appears between the meniscus and the RUC shock as in
Fig.~\ref{fig:profile}.
In the experiments, the leading UC shock is stable, while the RUC
shock becomes unstable and begins to finger as predicted\cite{M02}.
The dimensionless fingering wavenumber ($2 \pi l/\lambda $) is
$0.39\pm.02$ which is close to the most unstable wavenumber of $0.35$
predicted by linear theory\cite{M02}.  As the film evolves, a flat
region ($h_{\mathrm{RUC}}$) appears just behind the RUC shock.

\begin{figure}[h] 
\center{\parbox{3.2in}{\psfig{file=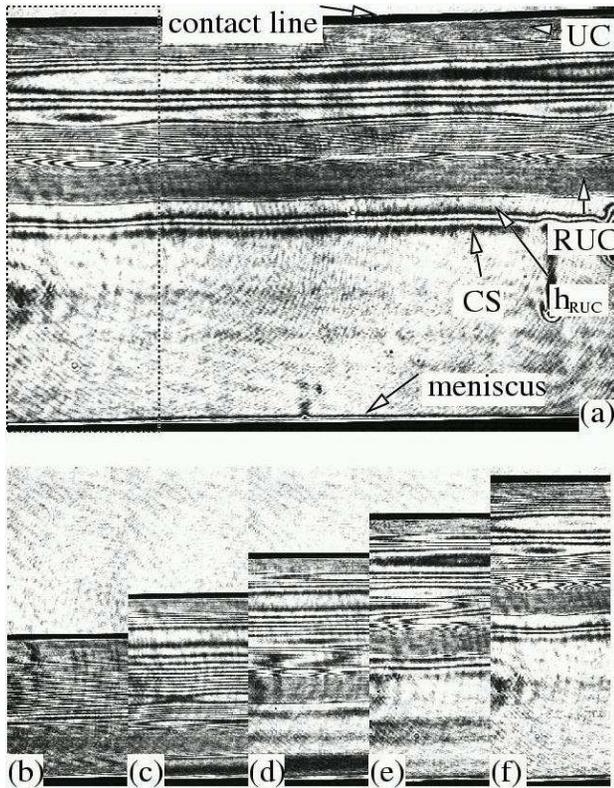,width=3.2in}}} 
\caption{(a) Interference fringes at 2400 sec. for 
$\alpha = 45^{\circ}$ and $\tau$ = 0.11 Pa.  The dimensions are as in 
Fig. ~\ref{fig:profile}. Sections (dashed region) are shown at 
(b) 0 sec, (c) 600 sec,
(d) 1200 sec, (e) 1800 sec, and (f) 2400 sec.  }
\label{fig:compressive} 
\end{figure}

The film thickness $h_{\mathrm{eq}}$ just above the meniscus is
determined by balancing surface tension gradient, gravity, and
curvature\cite{CC93}.  Thus, the inclination angle and the surface
tension gradient control $h_{\mathrm{eq}}$.  Here, we fix
$d\gamma/dx$ and we vary $\alpha$ to control $h_{\mathrm{eq}}$.
Decreasing $\alpha$, leads to increasing $h_{\mathrm{eq}}$ and
$h_{\mathrm{eq}} - h_{\mathrm{RUC}}$.  If $h_{\mathrm{eq}} >
h_{\mathrm{RUC}}$, the a flat region with a CS replaces the RW just
behind the $h_{\mathrm{RUC}}$ region as in Fig.~\ref{fig:compressive}.

\begin{figure}[h] 
\center{\parbox{3.2in}{\psfig{file=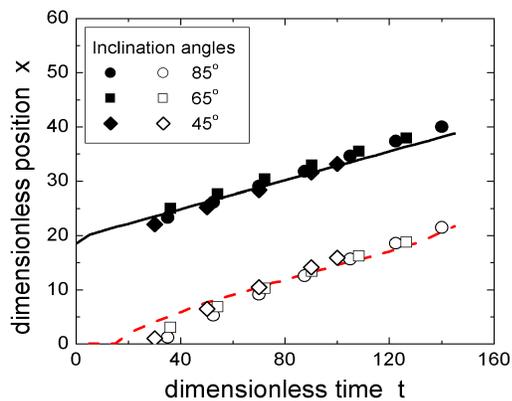,width=3.2in}}} 
\caption{The position of the contact line (solid symbols) and the position of 
the reverse undercompressive shock (open symbols) versus time. Theory
as predicted from the computation from Fig. 5 shown as solid and dashed lines.}
\label{fig:scaling} 
\end{figure}

The position of the leading UC shock is linear in time, while the RUC
shock shows transient nonlinear motion, consistent with the
theoretical model (see below). The leading UC shock is the same as the
one observed in the Cal experiments and the speed is the same as
in\cite{SC1-00}.  The RUC shock is always faster than the leading UC
shock, so that the size of the bump dwindles.  The RUC shock position
as a function of time for different inclination angles all collapse
when the position is rescaled by $l$, and time is rescaled by $T$, as
shown in Fig.~\ref{fig:scaling}.

To compare the experimental data with the model, we need boundary and
initial conditions.  Ahead of the contact line, we choose the simplest
boundary condition
 consistent with complete wetting: a precursor model with $h\to b>0$
as $x\to \infty$ \cite{BMFC98,fingerpaper,Troian}.  As shown 
previously\cite{BMFC98,SC1-00,MB99}, $b$ determines both the height
and speed of the leading UC shock.  This in turn determines the height
and speed of the
trailing reverse undercompressive wave.  We find that $b=0.005$ gives
the best approximation of the leading edge of the advancing film in these
experiments.  In the model, we take $x=0$ to be the position of the
edge of the meniscus.  For the lower boundary condition, we assume the
meniscus has just pinched off a film of thickness $h=h_{\rm eq}$ with
zero third derivative $h_{xxx}(0) = 0$, i.e.  that the meniscus
enforces a zero net curvature gradient on the bulk film region.  The
initial condition is the simplest approximation of a draining film,
that of constant film thickness $h_f$ for the range $0<x<x_f$, where
$x_f$ is given by the initial dipping and draining of the film.  The
parameters $x_f$ and $h_{\rm eq}$ can be measured precisely in the
experiment.  The value $h_f$ is chosen to produce a total volume of
fluid consistent with what is observed in the experiment.

We numerically integrate (\ref{non-dimen}) forward in time
and compare with the experimental data in Figs. 2 and 3.
In both cases $h_{\rm eq}$ and $x_f$ are determined from experimental
data and $h_f$ is chosen to match the volume of fluid observed
to pinch off in the experiment.  
The viscosity varies with temperature in the experiment.  We choose
a viscosity for the time rescaling of (\ref{non-dimen}) that is 
optimized for agreement with the data while consistent with the experiment.
\begin{figure}[h]
\center{\parbox{3.2in}{\psfig{file=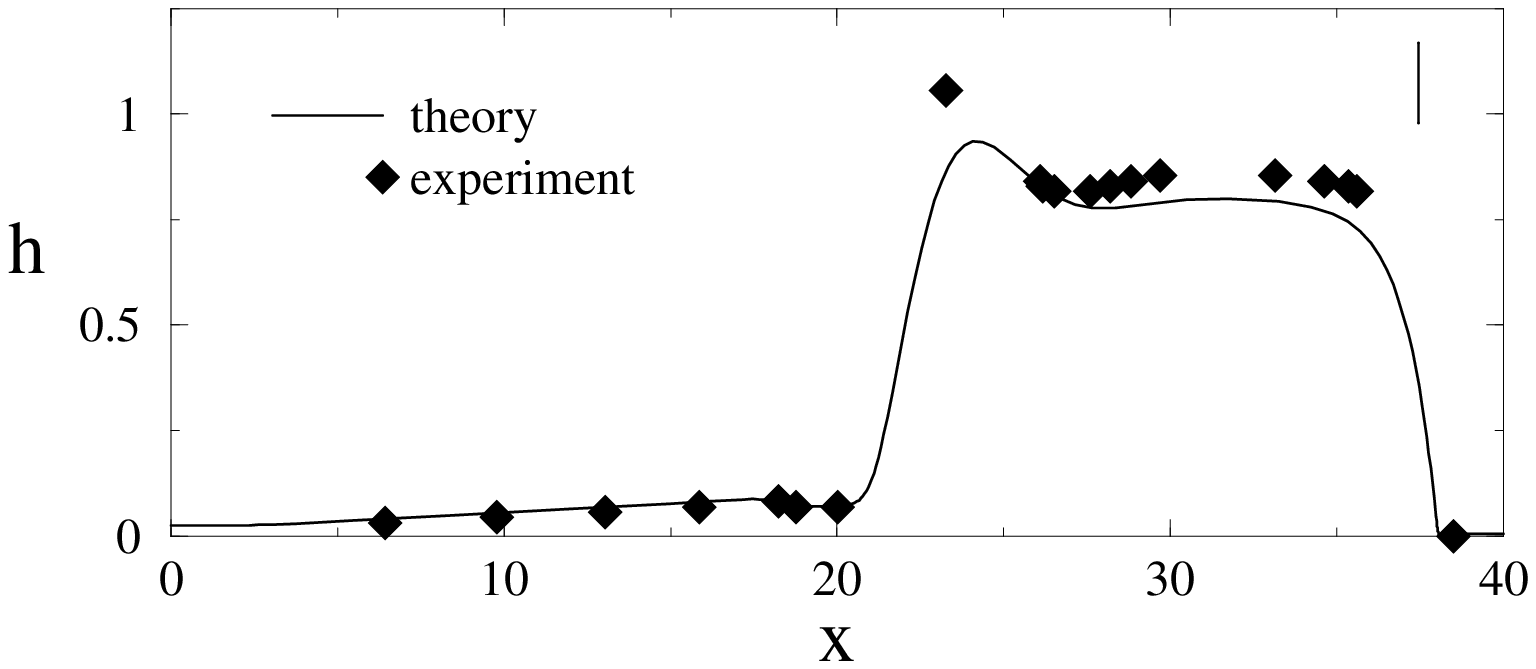,width=3.2in}}}
\center{\parbox{3.2in}{\psfig{file=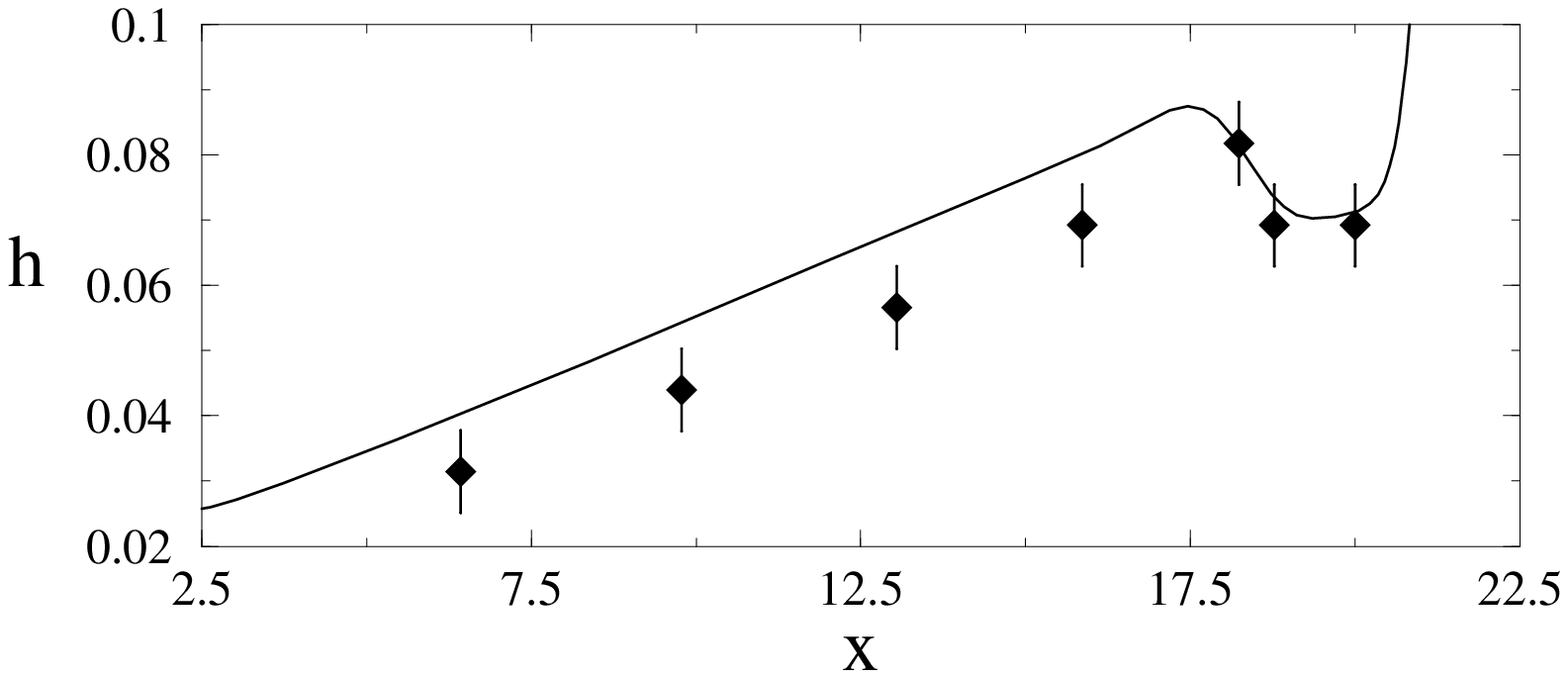,width=3.2in}}}
\caption{Comparison between PDE theory and experiment for data from Fig. 2a.  
Numerical parameters are $h_{\rm eq} = 0.025$, $h_f = 0.75$, $x_f = 18$.
The bottom figure shows a close up of the rarefaction wave. 
The bar in the top right
corner is the absolute error for $h>0.5$.  
The absolute error for smaller $h$ is shown on the bottom figure.  
The relative error
for $h>0.5$ equals the absolute error for the small $h$ values.}
\label{fig:compare3}
\end{figure}
Fig. 5 shows the theoretical thickness profile compared with 
the experiment from Fig. 2a, the dimensionless time is 140.    
The numerical solution evolves into a structure composed
of a leading UC shock, trailing RUC shock, and expanding RW, shown
closeup in the bottom portion of Fig. 5.
The leading UC shock speed and height are
determined by the precursor thickness
as in previous studies \cite{BMFC98,SC1-00}.  The RUC shock speed
and height are determined by the height of the leading UC shock,
and hence also determined by the precursor.  Fig. 5 shows a well
developed structure, however at early times the RUC experiences a period
of adjustment before it settles into a traveling wave shape, while
the leading UC shock forms almost immediately.  The trailing RW expands
between the meniscus region (set by the boundary condition $h_{\rm eq}$
in the model) and the RUC shock.  The trailing side of the RUC shock
has a thickness $h_{\rm RUC}$.  If $h_{\rm RUC}$ 
is greater than $h_{\rm eq}$, 
a RW forms between the RUC shock and the meniscus.  Otherwise
a CS forms to connect the meniscus pinch-off thickness $h_{\rm eq}$
to the thickness $h_{\rm RUC}$.  This CS is seen to slowly separate
from the RUC shock.
Fig. 6, which compares the PDE solution
with the data from Fig. 3a, in which $h_{\rm eq}>h_{\rm RUC}$.
A CS forms to the left
of the RUC shock, as shown close up in the bottom of Fig. 6.
At this $\alpha$ of $45^\circ$, we are close to this transition point,
and the experimental results are quite sensitive to changes 
in the initial draining film profile.
\begin{figure}[h]
\center{\parbox{3.2in}{\psfig{file=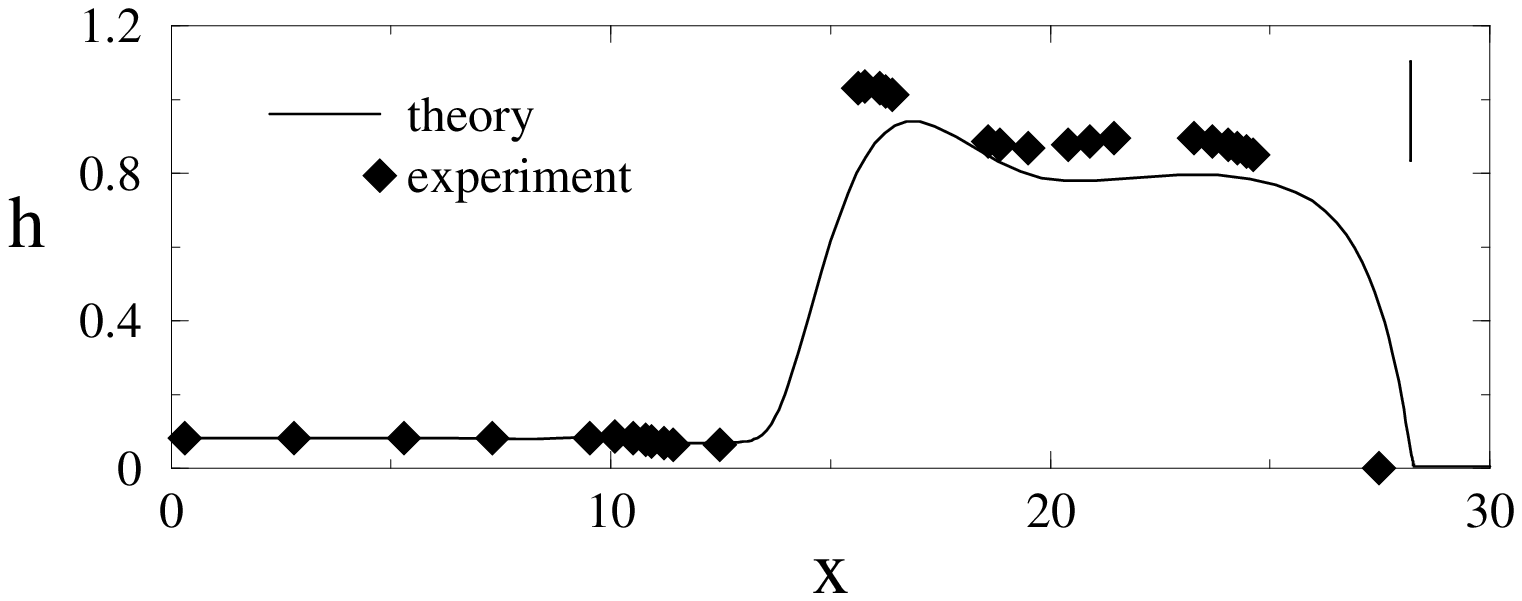,width=3.2in}}}
\center{\parbox{3.2in}{\psfig{file=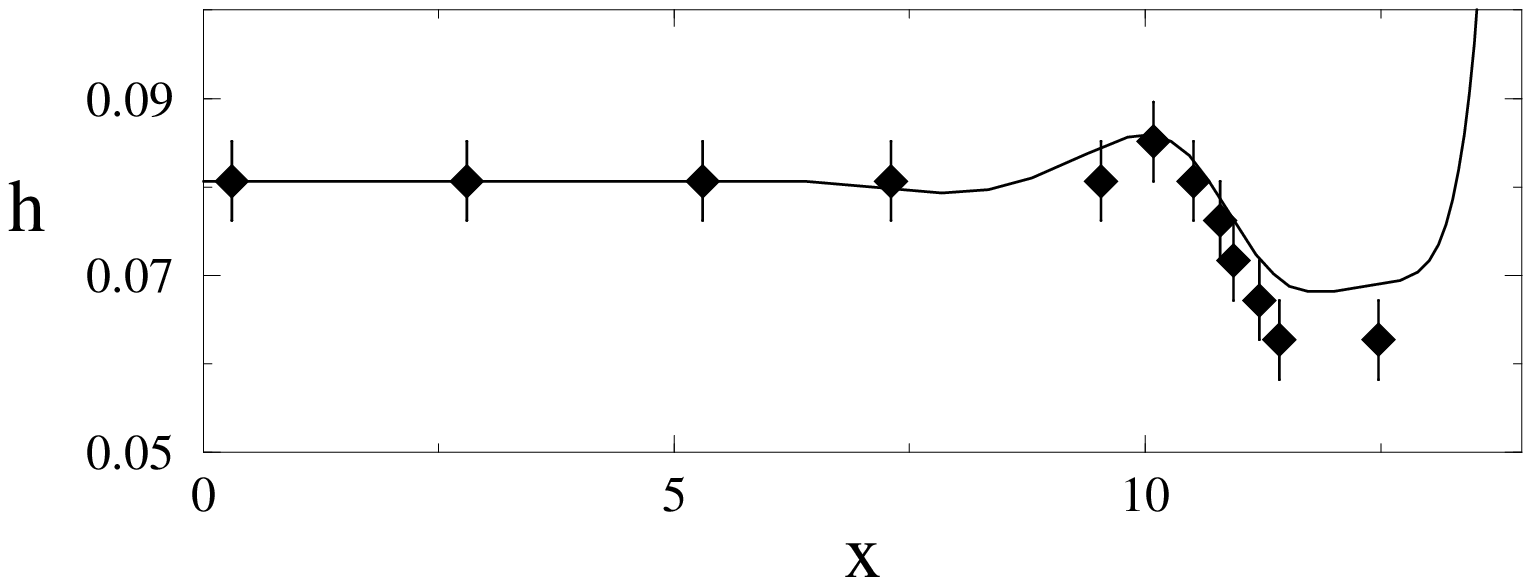,width=3.2in}}}
\caption{Comparison between PDE theory and experiment for data from Fig. 3a,
numerical parameters are $h_{\rm eq} = 0.081$, $h_f=0.75$, $x_f = 15.4$.
The error bars are as in Fig.~\ref{fig:compare3}.} 
\label{fig:compare4}
\end{figure}

The experiments presented show clear evidence of the new UC-RUC double
shock structure predicted by M\"unch.  
Our experimental results are compared with a lubrication model
in which the pinch-off dynamics of the meniscus are incorporated
into a boundary condition.  There is good agreement between theory and
experiment.
The dynamics between the RUC shock and the meniscus can result
in either a RW or a CS, depending on the pinch-off thickness
compared with the RUC film thickness $h_{\rm RUC}$.
Both situations are seen in the experiment and compared with theory.
Future studies should address
the nonlinear dynamics of the trailing RUC shock.
Also the effect of temperature on viscosity is not considered in the model here
but clearly plays a role in the experiment.   

We thank A. M\"unch for sharing an advanced copy of his manuscript
and A. M. Cazabat for helpful comments about the experiment.
This work is supported by NSF grant DMS-0074049 and ONR grant N000140110290.

\end{multicols}
\end{document}